# Glycine Zwitterion Stabilized by Four Water Molecules

**Byeong June Min**

*Department of Physics, Daegu University, Kyungsan 712-714, Korea*

We performed plane wave density functional theory calculations to survey the potential energy surface of neutral glycine (GlyNE) and its zwitterion (GlyZW) solvated by up to four water molecules. Our previous conformation study of Gly suggests inadequacy of the commonly used local basis function sets in dealing with a high-energy isomer, such as Gly. We find the potential energy surface of GlyNE and GlyZW smoother than usually was thought and without many local minima. Two water molecules can create a local minimum around GlyZW. With three water molecules, the energy difference between GlyNE and GlyZW is reduced to a mere 27 meV with an energy barrier of 115 meV from GlyNE. GlyZW becomes energetically more stable by 122 meV when solvated by four water molecules. Water molecules become catalysts in the tautomerization and sometimes engage in a switching transfer of a proton over the water bridge. Our results are consistent with experimental findings that the effective hydration number of Gly is 3 ~ 4.

## I. INTRODUCTION

Glycine (Gly) is the simplest amino acid with a side chain of single hydrogen atom. Amino acids in the presence of water favor zwitterionic form (GlyZW) in which the hydrogen in the carboxyl group is transferred to the amine group. This is the preliminary step for the formation of peptide bonds in proteins. As such, there have been many theoretical [1-5] and experimental studies [6, 7] on the neutral Gly (GlyNE), GlyZW, their dipole moment, and the transition pathway, with or without water molecules.

Considering the simplicity of Gly, we may be inclined to think there should remain nothing new to find out about Gly by this time. Not surprisingly, quite a many quantum chemistry calculations performed with a variety of local basis function sets and further corrections to include electron correlation energy unanimously predicted the same energy ordering for Gly conformers [1-5]. However, the predictions of the conformational energy ordering and the dipole moment were inconsistent with experimental findings. Theoretical dipole moment of about 1.0 Debye for gas phase Gly is inconsistent with 4.5 – 5.45 Debye [6, 7] from microwave spectroscopy. Even in the face of such disagreement, plane-wave density functional theory calculations have not been tried on Gly conformers. Molecular dynamics simulations have been performed on Gly solvated by 50 ~ 100 water molecules [8-10], using plane waves or mixed basis. In our previous plane wave density functional theory study on Gly isomers and conformers, we found a new energy ordering that is consistent with the experiments, against the unanimous predictions of existing local basis function results. We also verified that the local basis results are restored when the plane wave cutoff energy is reduced [11].

The erroneous energy ordering from local basis calculations has created a further confusion in the solvation study of Gly: the predicted most stable Gly conformer has a geometry that requires many water molecules to enable tautomerization, because the hydrogen on the carboxyl group is located at the farthest location from the possible tautomer site in the amine group. For that reason, researchers introduced effective medium and/or intricate water bridges. The preconception that Gly needs many water molecules for zwitterionization was so strong that the existing plane-wave (or mixed basis) studies consider Gly solvated by 50 ~ 100 water molecules, but not 1 ~ 4 water molecules.

Such conformational analysis also goes against intuition because GlyZW is then likely to have a small local minimum if the proton adsorbs on the far side of the nitrogen atom from the carboxyl group. Indeed, some local basis function studies find a local minimum around non-hydrated GlyZW in gas phase [12]. The absolute predominance of GlyNE over GlyZW in gas phase would also question such results. More significantly, the proton transfer path along the straight line connecting the nitrogen atom in the amine group and the oxygen atom in the carboxyl group could not prevail.

## II. CALCULATION

The calculations were performed using the ABINIT package [13] with periodic boundary condition and gamma point sampling. Norm-conserving pseudopotentials built by D. R. Hamann were used [14]. Plane wave energy cutoff of $1200\,\text{eV}$ is used. The cubic box dimension was chosen as 13.2 Å. The total-energy converges within $1\,\text{meV}$. Exchange correlation energy was described by the Perdew-Burke-Ernzerhof parameterization within the generalized gradient approximation (PBE-GGA) [15]. Semi-empirical treatment of the van der Waals interaction is used [16]. Self-consistency cycles were repeated until the difference of the total energy becomes smaller than $2.7\times10^{-6}\,\text{eV}$, twice in a row.

The nitrogen and two carbon atoms have been constrained to the z-plane. The internal reaction coordinate is the distance between the nitrogen atom and one of the hydrogen atoms. These two atoms are fixed, while all the atoms are fully relaxed. The system was relaxed until the average force on the atoms becomes smaller than $2.3 \times 10^{-3} \text{eV/Ang}$ by Broyden-Fletcher-Goldfarb-Shanno (BFGS) minimization scheme [17].

## III. RESULTS AND DISCUSSION

The geometries of hydrated GlyNE and GlyZW complex are shown in Figure 1 and the energy barrier between them in Figure 2.

GlyZW with zero or one water molecule is not stable and the potential energy surface does not have local minima around GlyZW. The total-energy was calculated as a function of $r_{N-H}$, fixing the positions of the N atom and the proton to prevent the recombination of the proton to the carboxyl group. The lowest energy configuration of GlyZW(H$_2$O) has the water molecule positioned between the amine group and the carboxyl group, to one side of the N-C-C center plane. Here, the tautomerization proceeds in a switching mechanism in which the water molecule gives away a proton to the amine group as it takes in a proton from the carboxyl group (Figure 1). For comparison, we define hypothetical GlyZW and GlyZW(H$_2$O) as the cases with $r_{N-H}=1.1$ Ang. The potential energy surface for GlyZW and GlyZW(H$_2$O) are monotonous, with a barrier of 762 meV and 476 meV with respect to GlyNE and GlyNE(H$_2$O), respectively. The energy barrier is already significantly reduced, almost by 40 %, when solvated by a single water molecule. It is indicative of the profound influence of water in the GlyNE and GlyZW energetics.

We find GlyZW(H$_2$O)$_2$ is in a local minimum, in agreement with Jensen *et al*. A loose water dimer runs parallel on one side of the N-C-C center plane and acts as a catalyst as the hydrogen atom transfers from the carboxyl group to the amine group. The position of the N atom and the hydrogen atom in the amine group toward the water dimer are fixed and the total energy is calculated as a function of the distance between them. Quite a small change in the N-H bond length achieves the tautomerization aided by the water dimer (Figure 1 and Figure 2). The GlyZW($H_2O$)$_2$ is 166 meV higher in energy than GlyNE($H_2O$)$_2$, with the energy barrier 189 meV from the GlyNE($H_2O$)$_2$ side (Figure 2). The potential well barrier is only 23 meV high from the GlyZW side, but it is deep enough to hold more than one vibrational state. The energy difference between GlyNE($H_2O$)$_2$ and GlyZW($H_2O$)$_2$ reported by Jensen *et al.* is 568 meV, exhibiting the ragged potential energy surface generated by local basis methods.

GlyNE($H_2O$)$_3$ and GlyZW($H_2O$)$_3$ are already very close in energy: GlyZW($H_2O$)$_3$ is only 27 meV higher than GlyNE($H_2O$)$_3$ and the energy barrier only 115 meV from Gly($H_2O$)$_3$. The water molecules line up along the center axis and to one side of the N-C-C center plane (Figure 1). The geometry is very close to that of GlyNE($H_2O$)$_2$ and GlyZW($H_2O$)$_2$ with an additional water molecule located near the carboxyl group. The three-water chain acts as a catalyst while the hydrogen atom of the carboxyl group migrates to the amine group, quite similarly as in GlyNE($H_2O$)$_2$ and GlyZW($H_2O$)$_2$.

GlyNE($H_2O$)$_4$ and GlyZW($H_2O$)$_4$ are accompanied by a three-water chain on one side of the N-C-C center plane and a water molecule on the opposite side. The tautomerization occurs by a switching mechanism involving the lone water molecule, while the three-water chain participates in the process as a catalyst. GlyZW($H_2O$)$_4$ is 122 meV lower in energy than GlyNE($H_2O$)$_4$, and the energy barrier between them is 235 meV from the GlyZW side. Experiments report a free energy difference of 0.31 eV and an enthalpy difference of 0.45 eV for GlyZW in aqueous solution [18].

We find the GlyNE and GlyZW potential energy surface from plane-wave calculations quite smooth and without many local minima, up to four water molecules. The number of water molecules needed to stabilize GlyZW turns out to be minimal: three water molecules make GlyNE and GlyZW almost isoenergetic and four water molecules stabilize GlyZW. The transition happens quicker than we could have guessed.

Figure 3 shows the binding energies as calculated by $E_b = E[Gly(H_2O)_n] - E[Gly] - E[(H_2O)_n]$. Here, $E[Gly]$ is the total energy of the lowest energy conformer of GlyNE, and $E[(H_2O)_n]$ the total energy of the water cluster. The binding energy for GlyZW(H$_2$O)$_4$ is 0.789 eV, already quite close to the experimental heat of solvation 0.83 eV [19]. The dipole moment of Gly(H$_2$O)$_n$ complex is shown in Figure 4. Dipole moment of GlyZW(H$_2$O)$_n$ for n = 2 ~ 4 are about 7 Debye and do not exhibit large variation. Microwave dielectric relaxation spectroscopy [20] reports 15.7 Debye and broadband dielectric relaxation spectroscopy 11.9 Debye [21]. Our results are quite smaller, but we must also consider that the experimental results on a system composed of two relaxing polar components are difficult to interpret.

We also note that Sato *et al.* [21] reports that Gly are bound by approximately 4.2 water molecules. Neutron scattering experiments by Kameda *et al.* [22] report that Gly is surrounded by 3 water molecules on the average. These results are compatible with our results in that the tautomerization is well advanced by the time when Gly is solvated by 4 water molecules.

## IV. CONCLUSION

Quite contrary to the widespread preconceptions, the tautomerization of Gly happens with only a few water molecules. GlyZW is already near isoenergetic when solvated by three water molecules, and more stable when solvated by four water molecules. Water molecules are very versatile and efficient in the tautomerization process. They sometimes act as catalysts, creating an environment for the H atom in the carboxyl group to make a direct migration to the amine group. They may also participate in the switching mechanism, in which they accept a proton from the carboxyl group and donate another proton to the amine group. Our results indicate that the tautomerization is well advanced by the time Gly is solvated by 4 water molecules, in agreement with the experiments [21, 22].

**ACKNOWLEDGMENTS**

This research was supported in part by the Daegu University Research Funds.

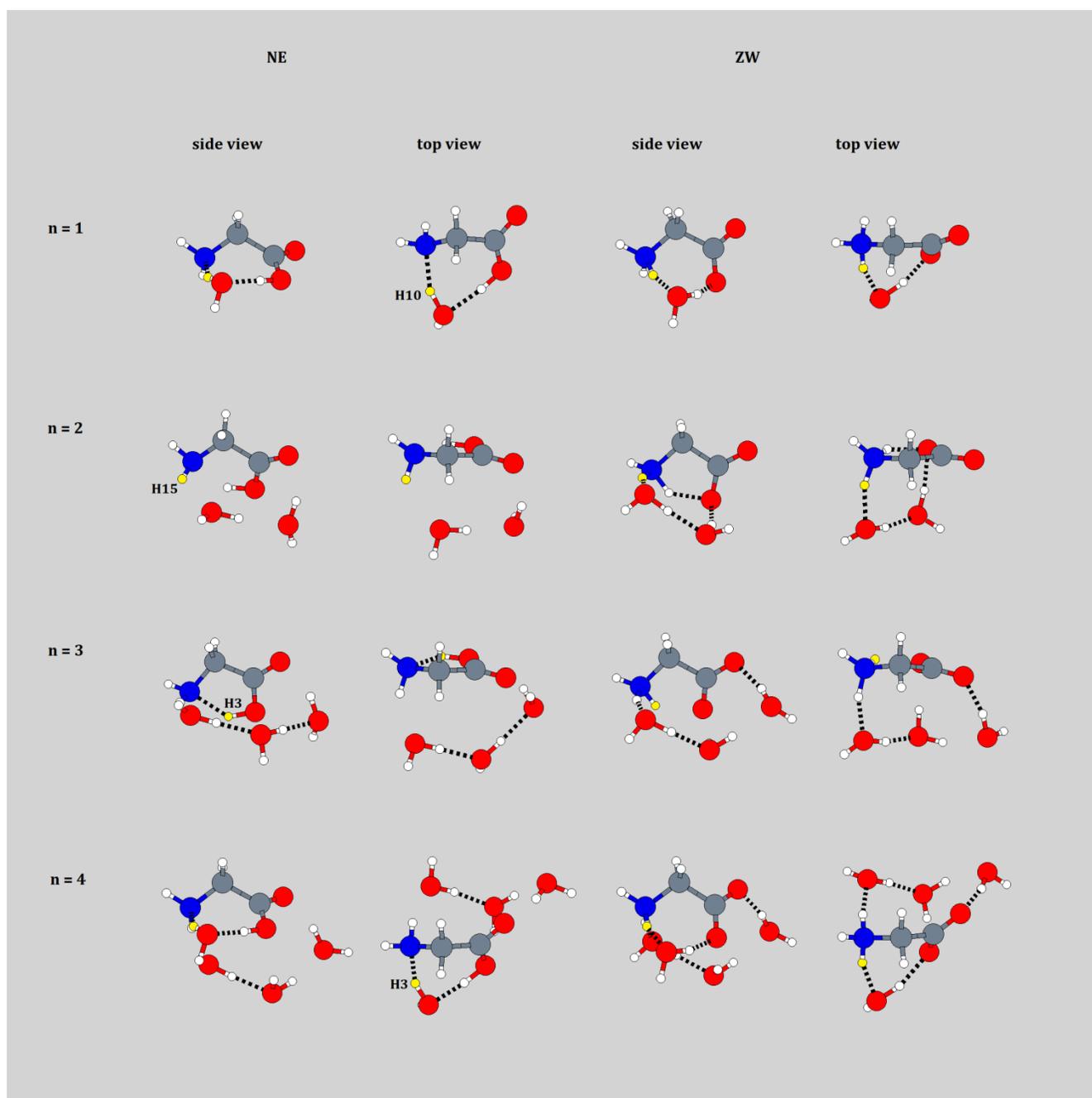

Fig. 1.

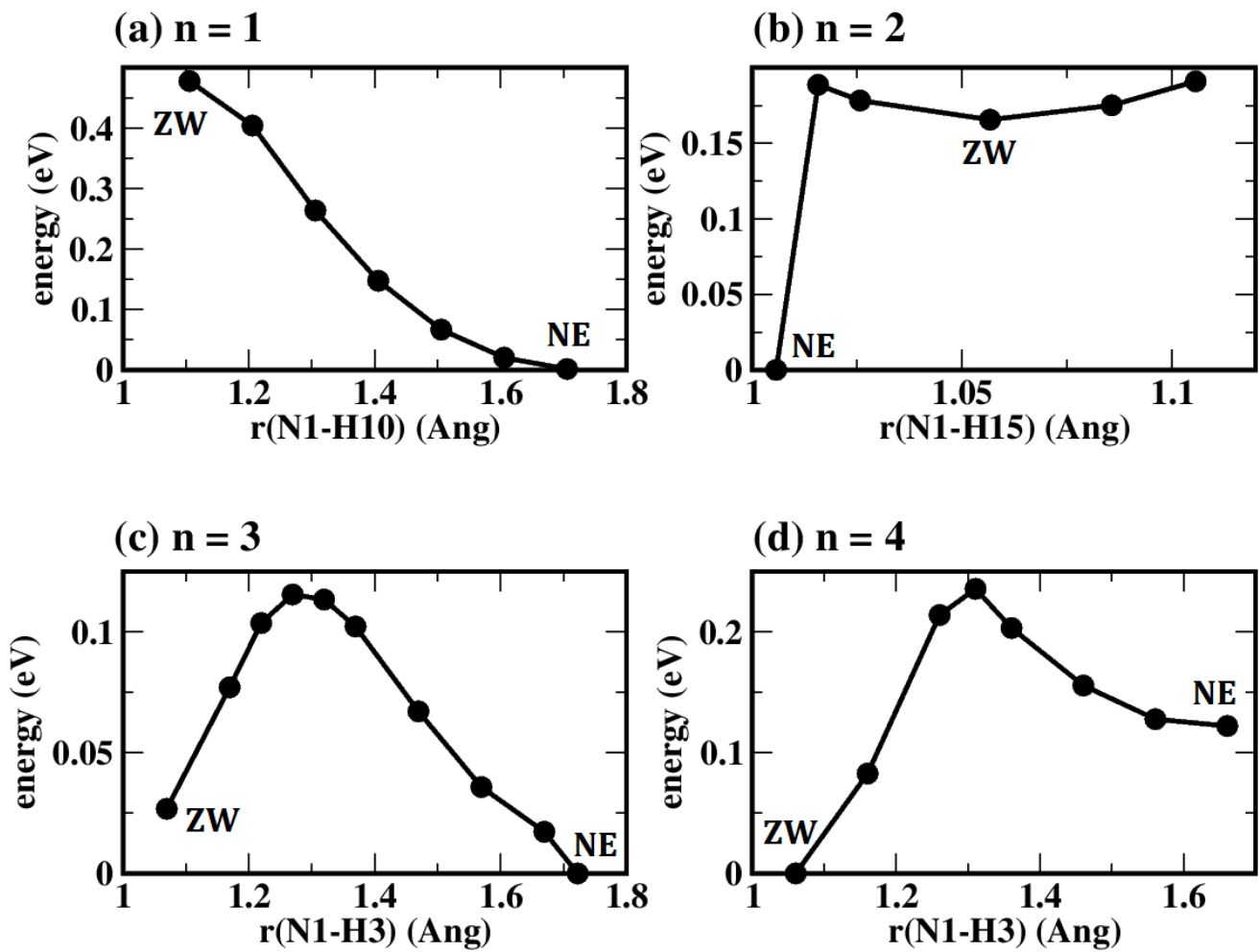

Fig. 2

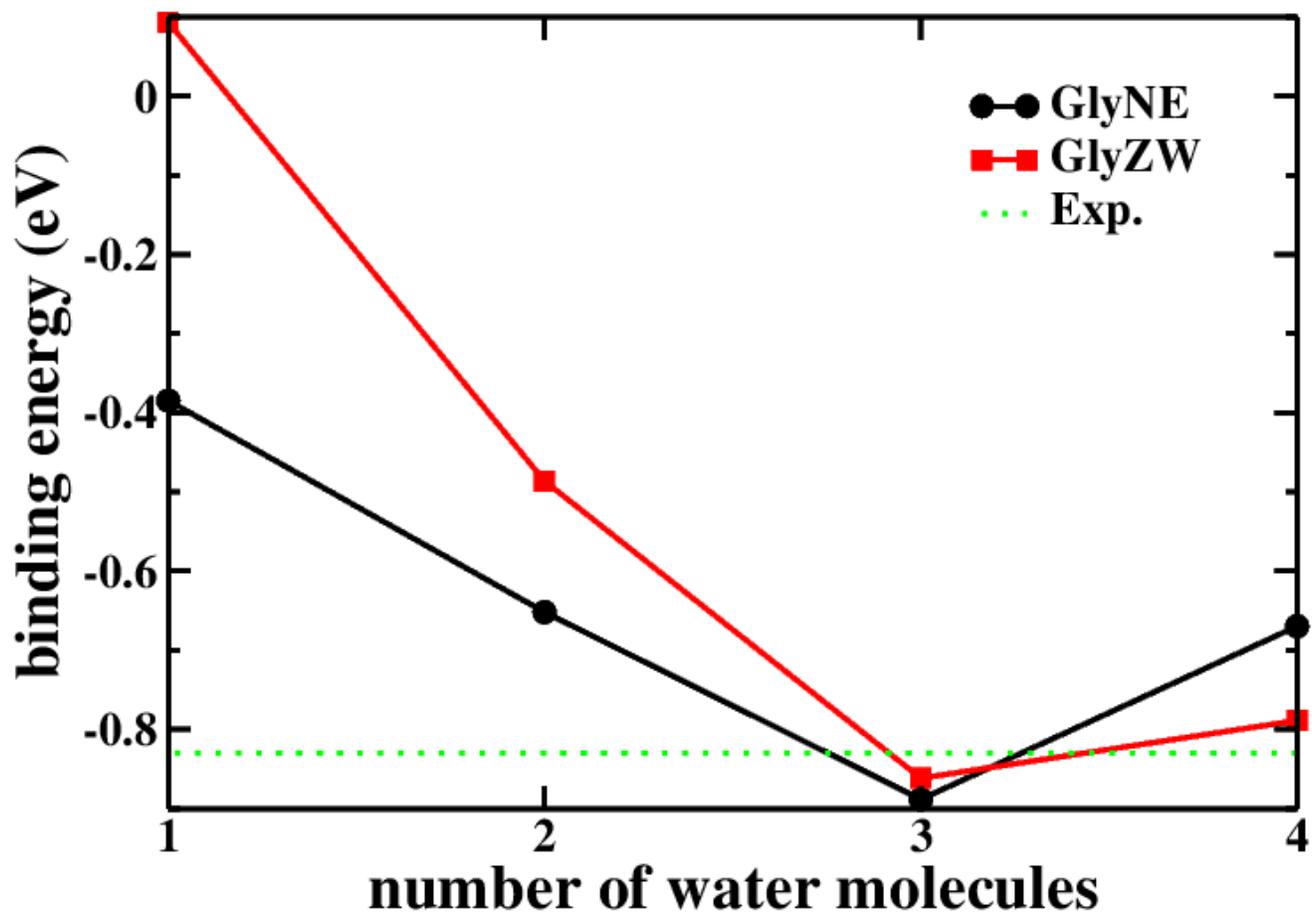

Fig. 3

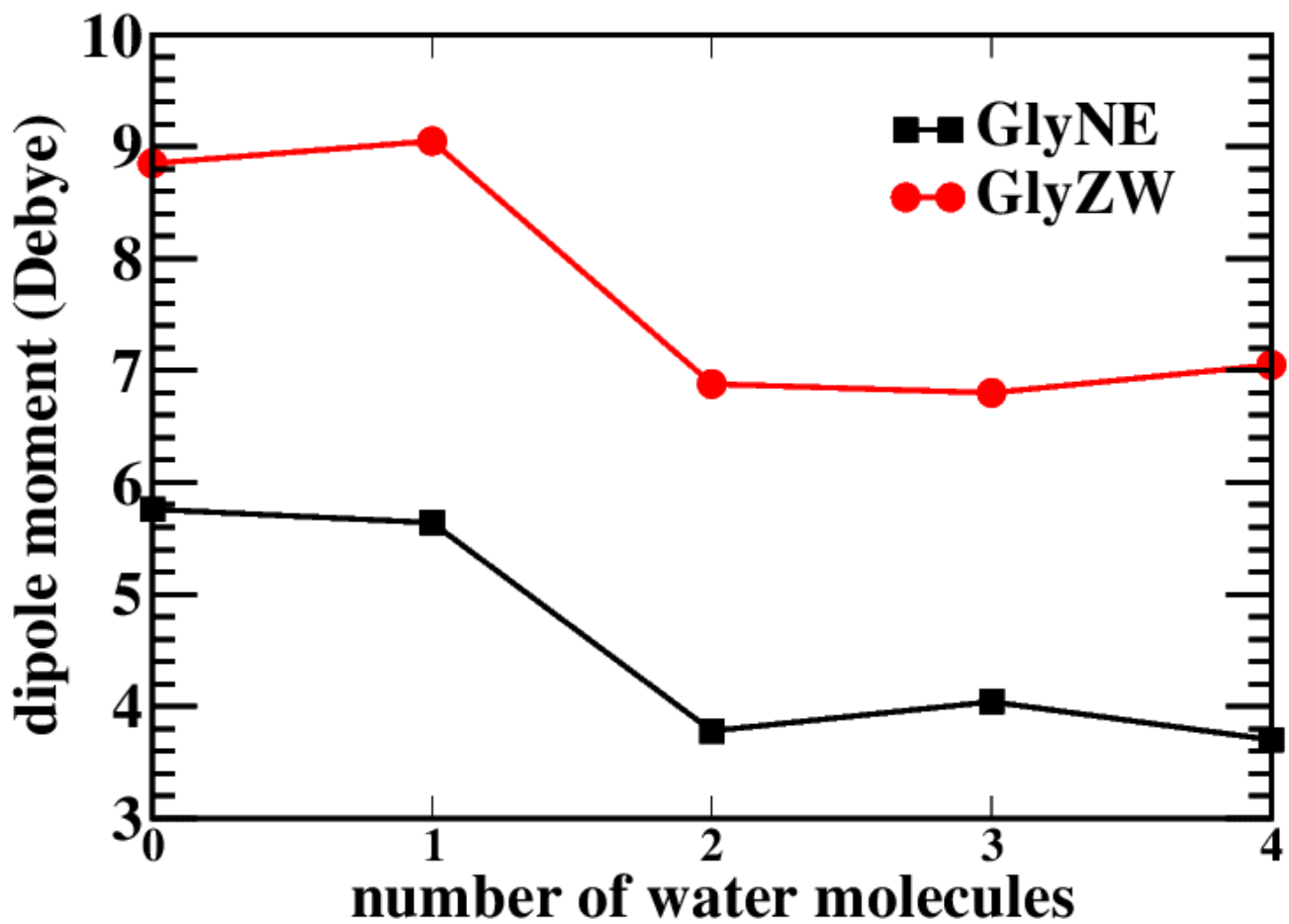

Figure 4

Figure Captions.

Fig. 1. Geometry of the neutral Gly (NE) and GlyZW (ZW) solvated by 1 ~ 4 water molecules. The distance between the N and H atom (filled with yellow) is used as the reaction coordinate. GlyZW solvated by single water molecule is hypothetical.

Fig. 2. The total energy as a function of the N-H distance (H atom shown in yellow in Figure 1).

Fig. 3. Binding energies of hydrated GlyNE and GlyZW complexes as calculated by $E_b = E[Gly(H_2O)_n] - E[Gly] - E[(H_2O)_n]$, compared with mass spectrometer experiment [19]. GlyZW solvated by single water molecule is hypothetical.

Fig. 4. Dipole moment of hydrated GlyNE and GlyZW complexes in Debye. GlyZW and GlyZW solvated by single water molecule is hypothetical.